\documentclass[10pt,a4paper]{article}
\usepackage{multirow}
\usepackage[dvips]{graphicx}

\usepackage[psamsfonts]{amssymb}
\usepackage{amsxtra}
\usepackage{threeparttable}
\usepackage{amsmath}
\usepackage{amssymb}

\usepackage{graphicx}

\usepackage{url}

\usepackage{color}

\usepackage{setspace}
\usepackage{float}

\usepackage[labelfont=bf,labelsep=period,justification=raggedright]{caption}

\makeatletter
\renewcommand{\@biblabel}[1]{\quad#1.}
\makeatother

\date{}

\begin{document}

\begin{flushleft}
{\Large
\textbf{Auditory Brain--Computer Interface Paradigm with Head Related Impulse Response--based Spatial Cues}\footnote{The final publication is available at IEEE Xplore \url{http://ieeexplore.ieee.org} and the copyright of the final version has been transferred to IEEE \copyright2013}

}
Chisaki Nakaizumi$^1$, Koichi Mori$^2$, Toshie Matsui$^1$, Shoji Makino$^1$,
and Tomasz M. Rutkowski$^{1,3,}$\footnote{The corresponding author. E-mail: \url{tomek@tara.tsukuba.ac.jp}}
\\
\bf{$^1$}Life Science Center of TARA, University of Tsukuba, Tsukuba, Japan\\
\bf{$^2$}Research Institute of National Rehabilitation Center for Persons with Disabilities, Tokorozawa, Japan\\
\bf{$^3$}RIKEN Brain Science Institute, Wako-shi, Japan\\
E-mail: \url{tomek@tara.tsukuba.ac.jp}\\
\url{http://bci-lab.info/}
\end{flushleft}

\section*{Abstract}

The aim of this study is to provide a comprehensive test of head related impulse response (HRIR) for an auditory spatial speller brain--computer interface (BCI) paradigm. The study is conducted with six users in an experimental set up based on five Japanese hiragana vowels. Auditory evoked potentials resulted with encouragingly good and stable ``aha--'' or P300--responses in real--world online BCI experiments. Our case study indicated that the auditory HRIR spatial sound reproduction paradigm could be a viable alternative to the established multi--loudspeaker surround sound BCI--speller applications, as far as healthy pilot study users are concerned.

\noindent{\bf Keywords:} Auditory BCI; P300; EEG; Brain Signal Processing.

\section{Introduction}
% no \IEEEPARstart

The majority of successful BCI applications rely on mental visual and motor imagery paradigms, which require long--term user training and good eyesight from the user~\cite{bciBOOKwolpaw}. Alternative solutions have been proposed recently to utilize spatial auditory~\cite{iwpash2009tomek} or somatosensory modalities~\cite{sssrBCI2006,tactileBCIwaiste2010,JNEtactileBCI2012,HiromuBCImeeting2013,tactileAUDIOvisualBCIcompare2013}. We propose to extend the previously published spatial auditory BCI (saBCI) paradigm~\cite{MoonJeongBCImeeting2013} by making use of a head--related--impulse--response (HRIR)~\cite{cipicHRTF} for virtual sound images spatialization with headphones--based sound reproduction. 
We test the concept with six BCI users, in a simple five Japanese hiragana (a, i, u, e, o) spatial speller task as previously proposed, and compared with auditory modality, using a vector--based--amplitude--panning (VBAP) approach~\cite{MoonJeongBCImeeting2013}. HRIR allows for more precise and fully spatial virtual sound images positioning utilizing even not own user's HRIR measurements~\cite{book:auditoryNEUROSCIENCE}. For the experiments reported in the paper we use a public domain \textsf{CIPIC HRTF Database} provided by the University of California, Davis (UC Davis)~\cite{cipicHRTF}.

The remainder of the paper is organized as follows. In the next section, the experimental setup and the HRIR--based saBCI paradigm are described, together with EEG signal acquisition, pre--processing and classification steps. Next, an analysis of evoked response potentials (ERP) and especially the so--called ``aha--' or $P300$ response latencies are described. Finally, classification and discussion of the HRIR--based saBCI paradigm information transfer rate (ITR) results conclude the paper, together with future research directions.

\section{Materials and Methods}

All of the experiments were performed in the Life Science Center of TARA, University of Tsukuba, Japan. Six volunteer users participated in the experiments. The average age of the users was $25$ years old (standard deviation $8.06$ years old,; one females and five males).
%The average age of the users was $25$ years old (standard deviation $8.06$ years old, average $25$ years old; one females and five males).
The psychophysical and online EEG BCI experiments were conducted in accordance with \emph{The World Medical Association Declaration of Helsinki - Ethical Principles for Medical Research Involving Human Subjects}. The experimental procedures were approved and designed in agreement with the ethical committee guidelines of the Faculty of Engineering, Information and Systems at University of Tsukuba, Japan.
The experiments were designed to reproduce the previously reported VBAP--based spatial auditory experiments~\cite{MoonJeongBCImeeting2013} this time with the more precise HRIR--based spatial auditory BCI stimulus reproduction to simplify the previously reported real sound sources generated with surround sound loudspeakers~\cite{iwpash2009tomek,bciSPATIALaudio2010}.

\subsection{Spatial Stimulus Sound Generation with HRIR and Reproduction to the User via Headphones}

The sound stimuli chosen for the subsequent psychophysical and saBCI experiment were the Japanese synthetic female voice utterances generated using a \emph{say} line--command of the \textsc{MacOS~X} computer operating system by Apple Inc. The synthetic female voice labeled as \emph{Kyoko} was chosen.
The five Japanese vowels \emph{a, i, u, e,} and \emph{o} were selected. Each stimulus was in a length range of $85-120$~ms. The generated synthetic sounds frequency spectra were within a limit of $0-11,000$~Hz. 

Next the the monaural stimulus sounds were spatialized by using the public domain \textsc{CIPIC HRTF Database}~\cite{cipicHRTF} and \textsc{Matlab} based signal processing functions. The HRIR model measured for the user~$\#124$ of the \textsc{CIPIC HRTF Database} was chosen.  Each Japanese vowel was set on horizontal plane with the same distance from the user's head at the azimuth locations of $-80^{\circ}$,$-40^{\circ}$, $0^{\circ}$, $40^{\circ}$, $80^{\circ}$ for the \emph{a, i, u, e, o} vowels respectively. In order to generate a stereo sound of a vowel placed at a spatial location at azimuth $\theta$ and elevation of $\phi$ the following procedure was applied. Let $h_{l,\theta,\phi}$ and $h_{r,\theta,\phi}$ be the minimum--phase impulse responses from the \textsc{CIPIC HRTF Database} measured at the chosen azimuth $\theta$ and the elevation $\phi$ at the left ($l$) and right ($r$) ears. The respective magnitude responses $H(f)$ obtained from Fourier transformed HRIR, the so--called head related transfer functions (HRTF), could be obtained as $|H_{l,\theta,\phi}(f)|$ and $|H_{r,\theta,\phi}(f)|$. The stereo spatial sound delivered via headphones to the left and right ears respectively could be constructed, in time domain using HRIR, as a two--dimensional signal composed of the left $x_l(t)$ and and right $x_r(t)$ headphone channels as follows,
\begin{eqnarray}
	x_l(t) = \sum^{n-1}_{\tau=0}h_{l,\theta,\phi}(\tau)s(t-\tau),\label{eq:HRIR1} \\ 
	x_r(t) = \sum^{n-1}_{\tau=0}h_{r,\theta,\phi}(\tau)s(t-\tau),\label{eq:HRIR2}
\end{eqnarray}
where $\tau$ denotes sample time delay and $n$ is the HRIR length as obtained from the \textsc{CIPIC HRTF Database}~\cite{cipicHRTF}.
The so optioned spatial acoustic stimuli were delivered to the left and right ears of the user using the ear--fitting portable headphones. 

\subsection{Psychophysical Experiment Protocol}

\begin{figure}[!h]
	\centering
	\vspace{4cm}
	\includegraphics[width = 0.9\linewidth,clip]{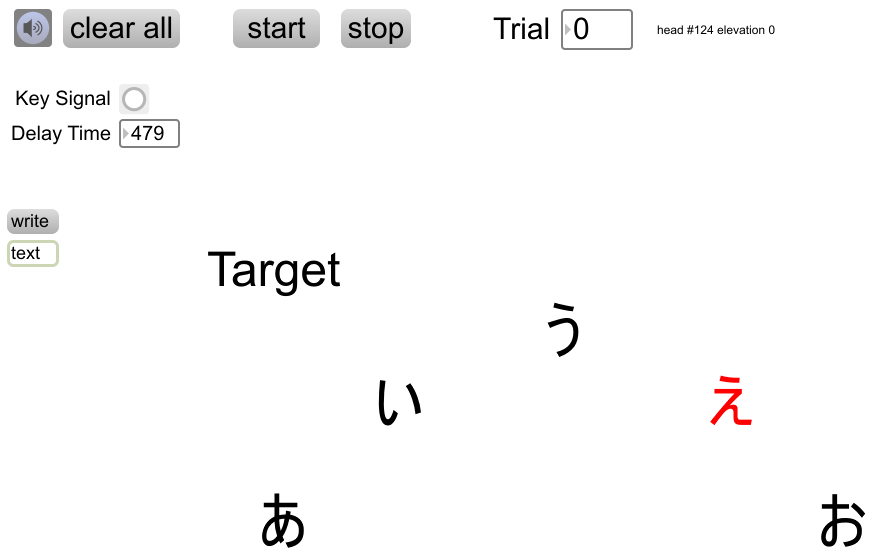}
	\vspace{0.5cm}
	\caption{The user instruction screen presented on a computer screen during the psychophysical and EEG experiments. The hiragana alphabet characters, displayed on the right in the figure, indicate the five Japanese vowels \emph{a, i, u,  e, o,} used in the experiments. The target stimulus in each trial (hiragana character \emph{``e''} in this particular instruction case in the figure) is emphasized by a red color. The interface buttons and experiment setup numbers depicted in the left and top areas of the user interface screen were used by the experimenter to setup and start each trial.}
	\label{fig:psycho-cap}
\end{figure}

The psychophysical experiments were conducted to investigate the response time and recognition accuracy by each user for the spatial Japanese vowel stimuli. The original monaural sound stimuli were spatialized using the HRIR--database filters as in equations~(\ref{eq:HRIR1})~and~(\ref{eq:HRIR2}) using a \textsc{MAX~6}~\cite{maxMSP} patcher developed by our team. An user instruction visual interface is depicted in Figure~\ref{fig:psycho-cap}. The Japanese hiragana character marked in red color, in the above figure, represented a current spatial \emph{target} which user should attend  and respond to using a computer keyboard as soon as possible after the perceived and recognized utterance. The other marked in black Japanese vowel stimuli, indicating \emph{non--targets}, were ignored as in a classical oddball paradigm~\cite{bciBOOKwolpaw}. The instruction which spatial \emph{target} stimulus to attend was delivered audio--visually, while during the following experimental oddball sequence only auditory modality was used.

The user was instructed in each trial to pay attention to the target stimulus vowel type and its spatial location, as well to respond by pressing a computer keyboard key as soon as the target stimulus was heard. The behavioral response times were collected with the same \textsc{MAX~6} patcher program used to display instructions (see Figure~\ref{fig:psycho-cap}) and to generate the HRIR spatial sounds.

Each single trial was comprised of randomly presented single \emph{target} and four \emph{non--target} vowel stimuli. An inter--stimulus--interval (ISI) was set for a one second in order to give the user time for coronet behavioral response. Each trial with presented five stimuli was separated with two seconds break.
We conducted $20$ trials for each vowel as the \emph{target} in the sequence. Finally we recorded responses of $100$ \emph{target} and $900$ \emph{non-target} stimuli in the psychophysical experiment. Results of the psychophysical experiments are discussed in the Section~\ref{seq:psychoRESULTS}.

\subsection{EEG Experiment Protocol}

In the EEG saBCI online experiment the brainwave signals were collected by a bio--signal amplifier system \textsf{g.USBamp} by g.tec Medical Engineering GmbH, Austria. The EEG signals were captured by sixteen active gel--based electrodes attached to the following head locations \emph{Cz, Pz, P3, P4, Cp5, Cp6, P1, P2, Poz, C1, C2, FC1, FC2,} and \emph{FCz} as in the extended $10/10$ international system~\cite{Jurcak20071600}. The ground electrode was attached on the forehead at the \emph{FPz} location, and the reference  on the left user's earlobe respectively. 

The spatial auditory stimuli were delivered, as in the preceding psychophysical experiments, using small headphones. 
The stimuli were generated using the same \textsc{MAX~6} patcher as in the psychophysical experiment (see Figure~\ref{fig:psycho-cap}) yet in this case triggers were produced by \textsc{an BCI2000}~\cite{BCI2000} EEG acquisition and processing environment used on our study. The triggers were communicated from \textsc{the BCI2000} to the \textsc{the MAX~6} patcher using an internet user datagram protocol (UDP).

\begin{figure}[!h]
	\centering
	%\vspace{-0.4cm}
	\includegraphics[clip,width=0.9\linewidth]{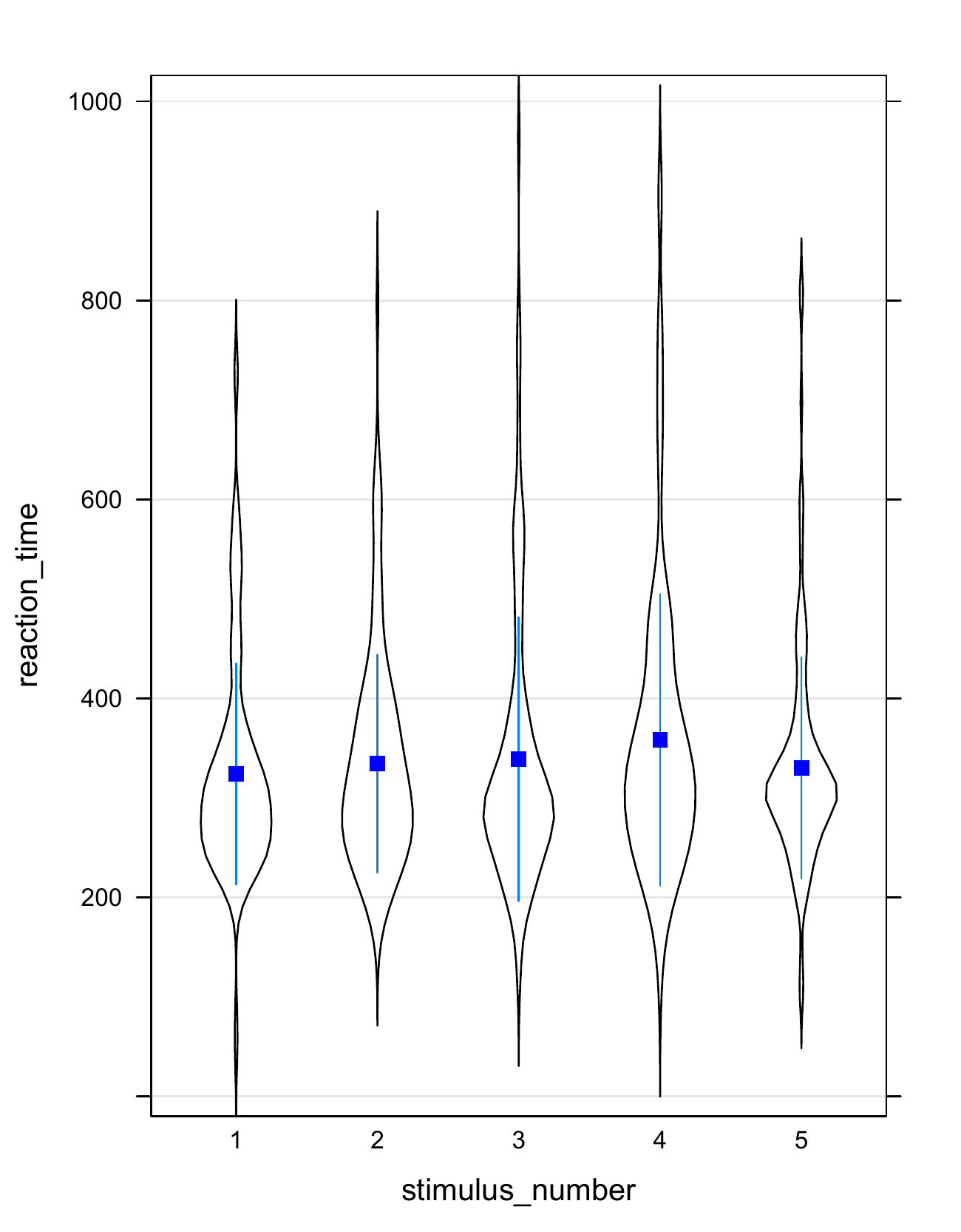}
	\caption{The psychophysical experiment response time probability distributions of all subjects together summarized in form of \emph{the violin--plots} with medians and interquartile ranges depicted. Each number at the horizontal axis represents the stimulus pattern $1\sim5$ representing the Japanese hiragana vowels \emph{a, i, u, e, o,} respectively. The horizontal axis represents the reaction (behavioral response) time delays in milliseconds.}
	\label{fig:psycho}
\end{figure}

\textsc{The BCI2000} environment's visual user interface was used to display the \emph{target} stimulus instructions. The online classification results were also presented using the same display. In a single Japanese hiragana vowel spelling session $10$ \emph{target} and $40$ \emph{non--target} stimuli were presented. Through the whole five vowels spelling session $50$ target and $200$ \emph{non--target} stimuli were delivered. The ISI in the EEG experiment was shorter comparing to the preceding psychophysical experiments and set in a range of $300\sim330$~ms (allowing for a random jitter within $30$~ms to break presentation's unnecessary rhythm), since the brain event related potentials (ERP), especially the ``aha--`` or P300--responses, had no jitter comparing to the behavioral keyboard presses (compare results in Figures~\ref{fig:psycho}~and~\ref{fig:EEGERP}). Each spatial auditory stimulus duration was set to $250$~ms. The EEG sampling rate was set to $512$~Hz and the notch filter to remove electric power lines interface of $50$~Hz was applied in a rejection band of $48-52$~Hz, according to the East Japan power stations specification. The high--pass filter was set at $0.1$~Hz and low--pass at $60$~Hz cutoff frequencies. Please note that the aim of this research is not noise and artifacts reduction~\cite{tomekJCSC2010}, but comprehensive test of HRIR usability for the novel saBCI paradigm.

The EEG signals were processed online by the \textsc{BCI2000}  application which used a stepwise linear discriminant analysis (SWLDA) classifier with features drawn from the $0-800$~ms ERP interval~\cite{caiCCpaper2013}. Results of the EEG saBCI paradigm experiments are discussed in the Section~\ref{seq:eegRESULTS}.

\section{Results}

This section presents and discusses results obtained from the psychophysical and EEG saBCI experiments conducted with six subjects as described in the previous sections.

\subsection{Psychophysical Experiment Results}\label{seq:psychoRESULTS}

\begin{figure}[!h]
	\centering
	\vspace{2cm}
	\includegraphics[clip,width=\linewidth]{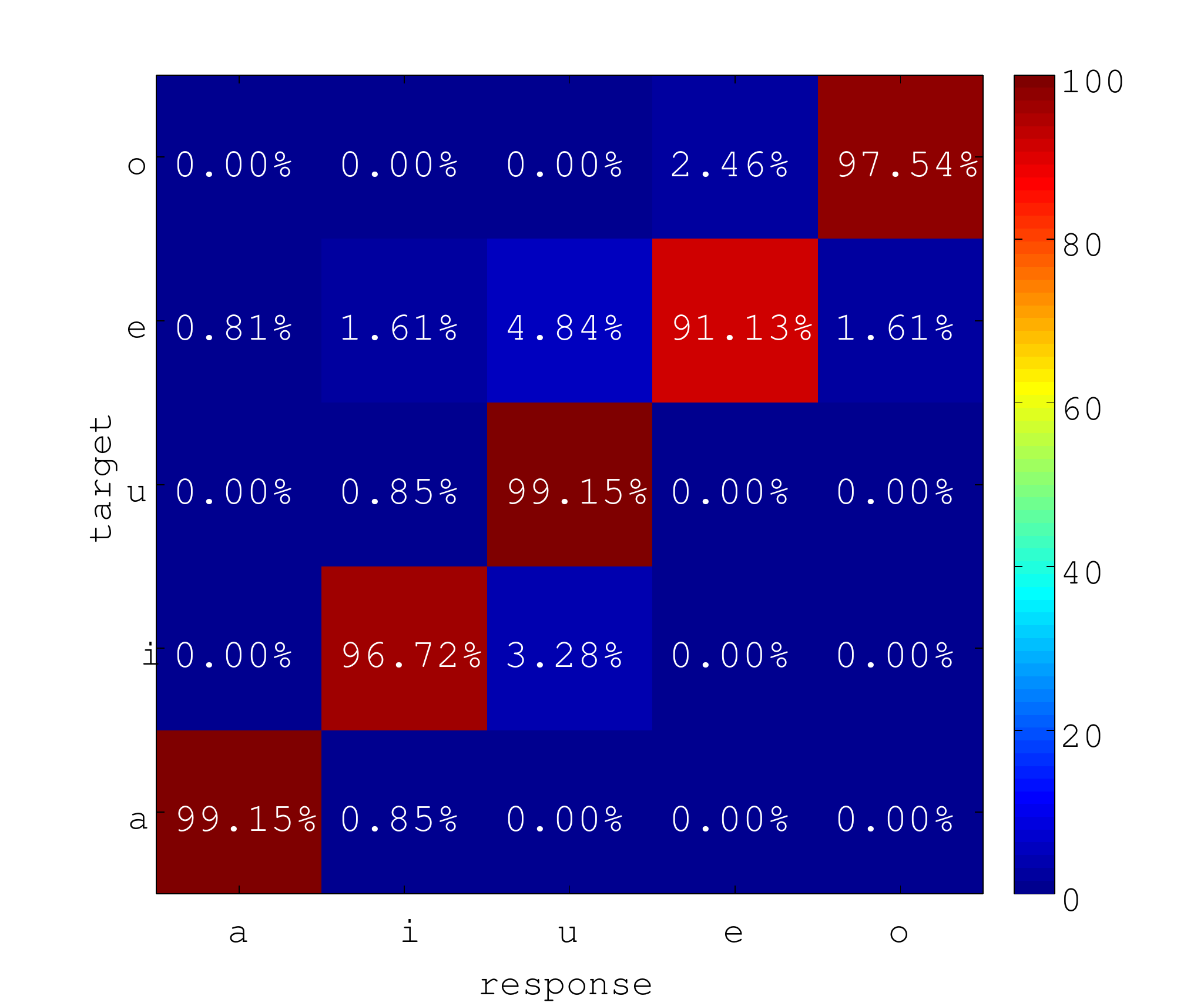}
	\vspace{0.1cm}
	\caption{User behavioral response accuracy in form a confusion matrix from the psychophysical experiment depicting averaged results. The numbers indicate the averaged accuracies which have been additionally depicted with color coding. The vertical axis indicates instructed \emph{targets} while the horizontal one the user responses.}
	\label{fig:psychoCMX}
\end{figure}

The psychophysical experiment user best response accuracy results are summarized in Table~\ref{tab:psycho}. All of the participants managed to reach perfect score at least once with $100\%$ accuracy which indicated the first confirmation of the HRIR--based spatial sound usability for the saBCI paradigm. Averaged responses are indicated in a form of confusion matrix depicted in Figure~\ref{fig:psychoCMX}. The confusion matrix has visualized on a diagonal the correct response accuracies (the majorities in our case) while on off--diagonals the mistakes. The averaged \emph{targets} were scored above $91\%$ accuracy. 

Figure~\ref{fig:psycho} presents analysis of response time delays in form of \emph{violin--plots} depicting distributions, median and interquartile ranges. \emph{Pairwise Wilcoxon rank sum tests} revealed no significant differences among medians of the response distributions further confirming the HRIR--based spatial stimuli for the saBCI experiments.

\subsection{EEG Experiment Results}\label{seq:eegRESULTS}

\begin{figure}[!h]
	\centering
	%\vspace{-0.6cm}
	\includegraphics[clip,width=0.8\linewidth]{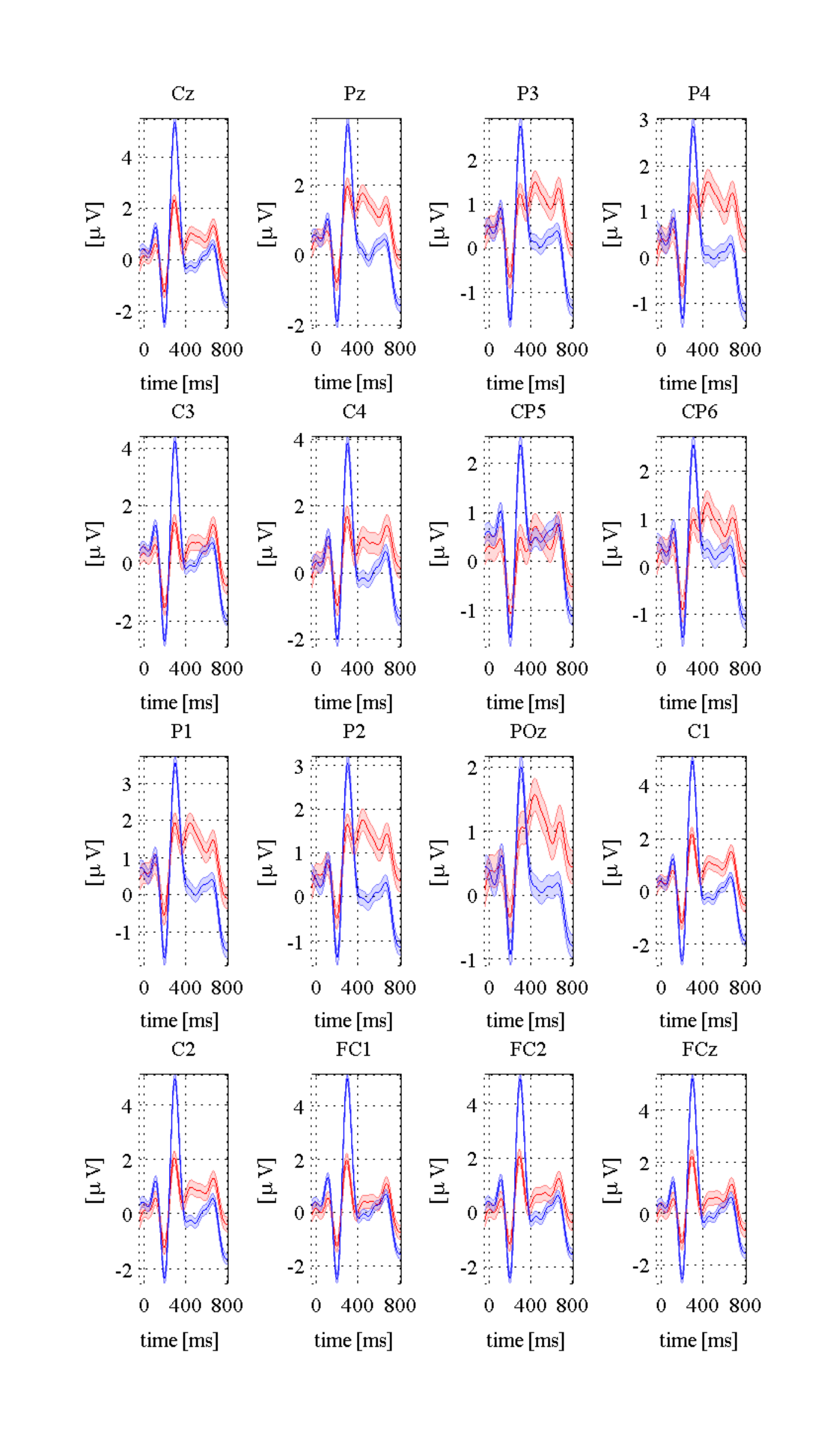}
	\vspace{-1.0cm}
	\caption{Grand mean averaged EEG ERP together with standard error bars for all participating users in the study plotted for each electrode separately. Red lines \emph{targets} and blue \emph{non--targets.}}
	\label{fig:EEGERP}
\end{figure}

\begin{figure}[!h]
	\centering
	%\vspace{-0.6cm}
	\includegraphics[clip,width=0.8\linewidth]{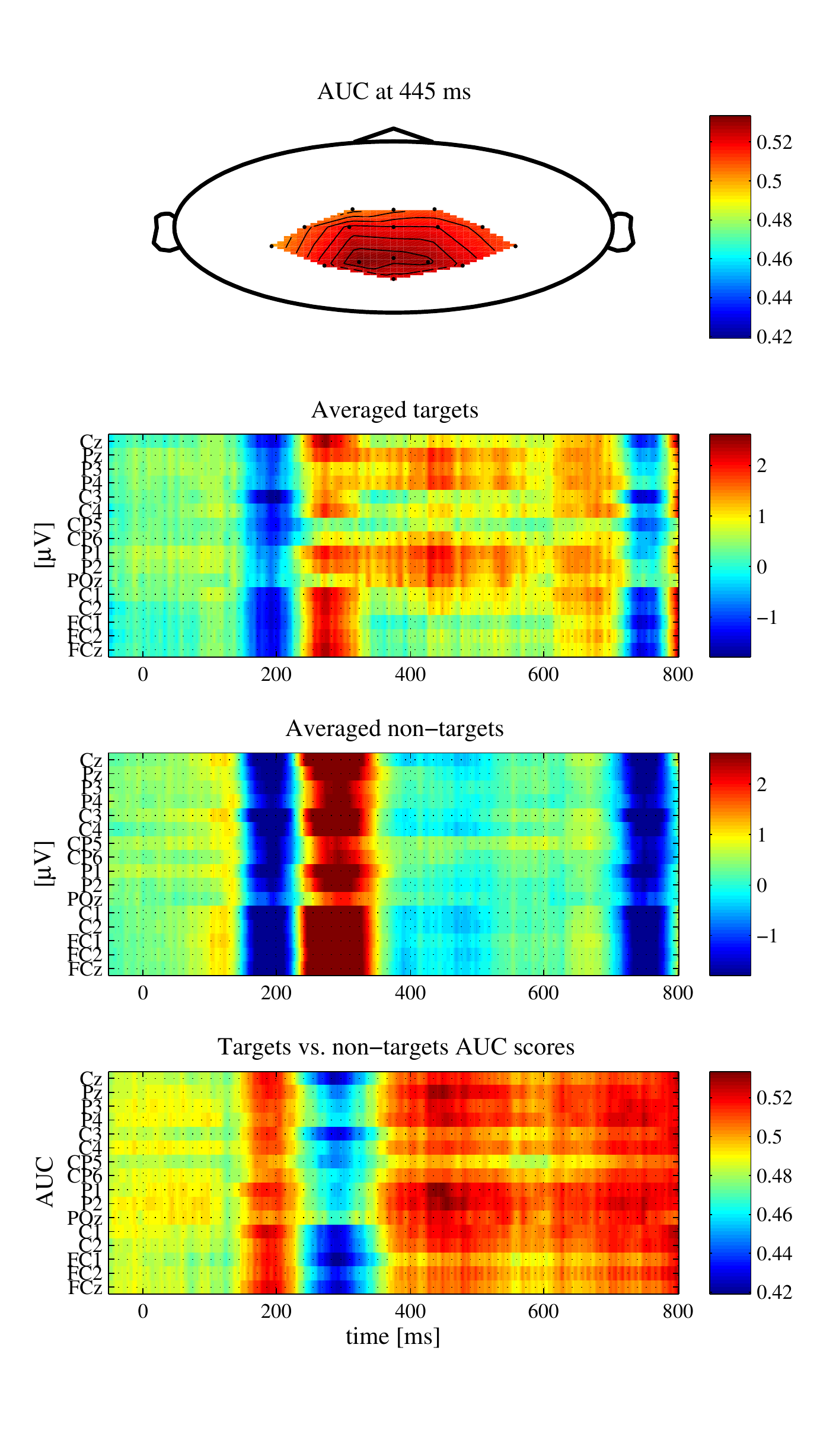}
	\vspace{-1.5cm}
	\caption{Grand mean averaged ERP and AUC scores leading to final classification results of the all participating users. The top panel presents the head topography at the maximum AUC score obtained from the bottom panel. The second panel presents averaged ERP responses to the target, while the third to \emph{non--target} stimuli respectively. The bottom panel visualizes the AUC analysis results of  \emph{target} versus \emph{non--target} response distribution differences.}
	\label{fig:EEGAUC}
\end{figure}

The results of the EEG experiment are summarized in Figures~\ref{fig:EEGERP}~and~\ref{fig:EEGAUC}, as well as in Tables~\ref{tab:results}~and~\ref{tab:ITR}. Table~\ref{tab:results} presents the online classification accuracies of the P300 responses as obtained with the SWLDA classifier. All the six users scored well above the chance level of $20\%$. There were two users who resulted with $100\%$ accuracies in online saBCI experiments. 

Figure~\ref{fig:EEGERP} presents the averaged ERP brain responses for all electrodes separately together with standard error bars to \emph{target} (red lines) and \emph{non--target} (blue lines) stimuli. The obvious P300 responses could be clearly seen in the $400\sim700$~ms latencies.

Figure~\ref{fig:EEGAUC} summarizes the results as a scalp topography in the top panel with maximum areas under the curve (AUC)~\cite{bciSPATIALaudio2010} values for  \emph{target} vs. \emph{non--target} latencies. 
The topographic plot also reflects the EEG electrodes positions used in the experiments. Although most of the electrodes produced significant P300 potential differences (see also Figure~\ref{fig:EEGERP}), the parietal cortex (the back of the head) resulted with more pronounced differences. The second panel from the top in the Figure~\ref{fig:EEGAUC} indicates the averaged ERP responses to the \emph{target}, while the third panel to the \emph{non--target} stimuli respectively. The bottom panel indicates the AUC of \emph{target} versus \emph{non-target} responses clearly confirming the $400\sim700$~ms latencies usability for the subsequent classification.

The presented EEG ERP results confirmed that the HRIR--based spatial auditory stimuli evoked the P300 response possible to classify in the online saBCI experiments with $10$--trails averaging.

Table~\ref{tab:ITR} reports the obtained ITR scores in a range of $4.80\sim9.29$~bit/min. The average ITR for the six participants in our study was $6.29$~bit/min. The ITR was calculated as follows\cite{bciSPATIALaudio2010}:
\begin{eqnarray}
	ITR &=& V\cdot R  \\
 	R &=& \log_2 N + P\cdot \log_2 P +\\
   	&+& (1-P)\cdot \log_2\left(\frac{1-P}{N-1}\right) \nonumber
\end{eqnarray}
where $V$ was the classification speed in selections/minute ($4$~selections/minute for this study); $R$ represented the number of bits/selection; $N$ was the number of classes ($5$ in this study); $P$ was the classifier accuracy as in the Table~\ref{tab:results}.

\section{Conclusions}

We conducted a series of psychophysical and online EEG saBCI paradigm experiments in order to evaluate the HRIR--based spatial sound generation system usability. In the psychophysical experiment, the HRIR--based spatial stimuli resulted with close to perfect accuracies.
The online saBCI EEG experiment results also show that the HRIR--based spatial sound reproduction method allows for even $100\%$ accuracy as shown with two subject results.
This result suggests a great potential in simple headphone--based HRIR spatial sound generation for the saBCI.
The presented results confirmed the hypothesis of the possibility to simplify spatial sound--based saBCI paradigms. 

This saBCI scored with high accuracy results without bulky multiple spatially distributed loudspeakers to generate stimuli. Nevertheless, current pilot study is not yet ready to compete with faster visual BCI spellers, for example. Furthermore, it is necessary to improve ITR for comfortable spelling. We plan to continue research with larger number of sound stimuli, with shorter ISI, and with more complex spatial sound patterns.

\section*{Author Contributions}

Designed and performed the EEG experiments: CN, TMR. Analyzed the data: CN, TMR. Conceived the concept of the HRIR--based saBCI paradigm: TMR, KM. Supported the project: SM, KM, TM. Wrote the paper: CN, TMR. 

% conference papers do not normally have an appendix
% use section* for acknowledgement
\section*{Acknowledgment}

This research was supported in part by the Strategic Information and Communications R\&D Promotion Program (SCOPE) no. 121803027 of The Ministry of Internal Affairs and Communication in Japan.

%\bibliographystyle{vancouver}
%\bibliography{chisaki}

\newpage

\section*{Tables}

\begin{table}[!h]
\begin{center}
%\begin{threeparttable}
\caption{Psychophysical experiment results (note, this is not a binary accuracy case yet the one with a theoretical chance level of $20\%$) in spatial auditory spelling task}\label{tab:psycho}
\begin{tabular}{| c | c |}
\hline 
User number 	& The best psychophysical accuracy \\
\hline \hline
$\#1$			& $100\%$ \\
$\#2$			& $100\%$ \\
$\#3$			& $100\%$ \\
$\#4$			& $100\%$ \\
$\#5$			& $100\%$ \\
$\#6$			& $100\%$ \\
\hline \hline
{\bf Average:}	& $\bf 100\%$ \\
\hline
\end{tabular}
%\end{threeparttable}
\end{center}
\end{table}

%\newpage

\begin{table}[!h]
\begin{center}
%\begin{threeparttable}
\caption{Ten averaged trials based BCI accuracy (note, this is not binary P300 classification result but resulting spelling result with a theoretical chance level of $20\%$) in spatial auditory BCI spelling task using the classical SWLDA classifier}\label{tab:results}
\begin{tabular}{| c | c |}
\hline
User number 	& Online BCI experiment SWLDA accuracy \\
\hline \hline
$\#1$			& $80\%$ \\
$\#2$			& $80\%$ \\
$\#3$			& $100\%$ \\
$\#4$			& $100\%$ \\
$\#5$			& $80\%$ \\
$\#6$			& $80\%$ \\
\hline \hline
{\bf Average:}	& $\bf 86.67\%$ \\
\hline
\end{tabular}
%\end{threeparttable}
\end{center}
\end{table}

%\newpage

\begin{table}[!h]
\begin{center}
%\begin{threeparttable}
\caption{Ten averaged trials based spelling accuracy (see Table~\ref{tab:results}) ITR results}\label{tab:ITR}
\begin{tabular}{|c|c|}
\hline
User number 	& ITR scores \\
\hline \hline
$\#1$			& $4.80$~bit/min \\
$\#2$			& $4.80$~bit/min \\
$\#3$			& $9.29$~bit/min \\
$\#4$			& $9.29$~bit/min \\
$\#5$			& $4.80$~bit/min \\
$\#6$			& $4.80$~bit/min \\
\hline \hline
{\bf Average:}	& $\bf 6.29$~bit/min \\
\hline
\end{tabular}
%\end{threeparttable}
\end{center}
\end{table}

\newpage

\section*{Figures}

\newpage

\newpage

\newpage

\newpage

\end{document}